\documentstyle[preprint,aps]{revtex}
\begin{document}
\draft
\title{Range of the t--J model parameters for CuO$_{2}$ plane:
experimental data constraints.}
\author{V. I. Belinicher, A. L. Chernyshev}
\address{Institute of Semiconductor Physics, 630090, Novosibirsk, Russia}
\author{L. V. Popovich}
\address{Institute of Inorganic Chemistry, 630090,  Novosibirsk, Russia}
\date{\today}
\maketitle
\begin{abstract}
The t-J model effective hopping
integral is determined from the three-band Hubbard model for the charge
carriers in CuO$_{2}$ plane.  For this purpose  the values of the
superexchange  constant $J$  and the charge-transfer gap $E_{gap}$ are
calculated in the framework  of the three-band model. Fitting values of $J$
and $E_{gap}$ to the experimental data allows to narrow the uncertainty
region of the three-band model parameters.  As a result, the  $t/J$ ratio of
the t-J model is fixed in the range $2.4 \div 2.7$ for holes and $2.5
\div 3.0$ for electrons.  Formation of the Frenkel  exciton is justified
and the main features of the charge-transfer spectrum are correctly
described in the framework of this approach.

\vskip 0.5cm \noindent
\end{abstract}
\pacs{75.10.Jm, 75.30.Ds}
\narrowtext
%************************************************************************
\section{Introduction}
\label{sec:level1}

%************************************************************************
A large amount of works dedicated to high-$T_{c}$ superconductors
agree that an appropriate    electronic model which contains all
essential orbitals is the three-band Hubbard model \cite{Em1,Va,Lok}.
Some other works developing a multiband approach \cite{Hy2,Al1,Esk0,Esk1}
also support this belief.

It is also widely accepted that the low-energy physics of insulating
compounds can be described by the Heisenberg Hamiltonian.  According to
the earlier work by Zhang and Rice \cite{Zh1}  and latter
studies \cite{Zh2}-\cite{Be3} lightly doped systems are described by the
simple t-J model.  Different techniques provide different
exactness of the three-band model to t-J model mapping and generate a
wide type of generalizations \cite{Yu1,Al2,Be3}.

In our previous works \cite{Be1,Be2,Be3} a consistent
low-energy reduction of the three-band model to the generalized t-J
model in the realistic range of parameters has been performed.  It
has been shown \cite{Be3} that the second-order corrections to the
local energy of the carrier and its hopping integral are small ($<5\%$).
The role of the next-nearest-neighbor terms   has also been discussed.

It is commonly believed that the 90\% or even higher accuracy  of the
t-J model as the low-energy electronic model for high-$T_{c}$
superconductors justifies its wider study \cite{Da1,Lu2,Su1}
and remains this model the main pretender in describing of the
superconductivity in cuprates \cite{Su2,Su3,Ed1}. Recent angle-resolved
photoemission experiments \cite{De1} can be interpreted as a direct
support of some t-J model properties \cite{Da2}.  Some other
anomalous behaviors of systems in normal state as well as
superconductivity itself seem to be described in the framework of
the t-J model \cite{Su2,Su3,Ed1}, \cite{Da2}.

The reasonable question from this point of view is: what  is the role
of either the three-band or more complex first-principles models?  There
are several answers: (i) calculating parameters for the t-J model
for real systems; (ii) giving  insight into the experiments
including not only the simple t-J model degrees of freedom.

In this paper  we mainly address the question of the t-J model
parameters. Superexchange constant $J$ for the t-J model is directly
measured \cite{Ro,Hay1}. Hence, the parameter to be determined is the
effective nearest-neighbor hopping integral.  The problem of its calculation
is not connected with the accuracy of low-energy mapping from the
three-band model, which is always very high,  but results from
uncertainty in the three-band model parameters.  The three-band model in
conventional formulation contains as inner parameters two on-site and
one inter-site Coulomb repulsions, two hopping integrals, and split
of the levels \cite{Em1} which are not directly measurable.  Some of
them are fairly bad determined.  This makes the calculation of
the hopping integral for real systems questionable and even
controversial.

We develop an obvious idea of fixing the three-band model parameters by
using experimental data.  This idea has already been exploited in the cluster
calculations for spectroscopic data \cite{Esk0}  and for the superexchange
$J$ in our previous work \cite{Be3}. Now, on the basis of better
understanding of the low-energy model of the electronic system and
magnetic polaron nature of the t-J model carriers \cite{Su1,Au1,Ed2},
we calculate quite accurately  the charge-transfer gap.
Selfconsistent calculation provides a narrow range of possible values
of $t/J$. Also, an excitonic feature of the charge-transfer
spectrum is obtained in agreement with resent experiments \cite{Fa1}.

Paper organized as follows.  In Sec. II we discuss the low-energy
limit of the three-band model and experimentally observable
quantities.  In Sec. III the calculation of the superexchange constant
$J$ and the charge-transfer gap $E_{gap}$ is produced.  In Sec. IV we
discuss the dependence $J$ and $E_{gap}$ on the parameters of the
three-band model and determine the range of the hopping parameter $t$
for electrons and holes. In Sec. V the properties of the excitonic
state is considered. Sec. VI presents our results and discussions.  The
technical aspects of work are given in Appendix.

%***************************************************************************
\section {The low-energy limit of the three-band model and observable
 quantities}

%***************************************************************************
Previously, it was suggested that the three-band Hubbard model is an
appropriate starting point for describing the electronic structure of
CuO$_{2}$ planes \cite{Em1,Va}.  The Cu $\ d_{x^{2}-y^{2}}$ orbital and
$p\sigma (x,y)$ orbitals are strongly hybridizated.  These orbitals are
explicitly treated in the three-band model with the justifiable assumption
that other orbitals does not directly participate in the low-energy dynamics.
The full Hamiltonian of the model is defined by \cite{Em1}
\begin{eqnarray}
\label{1}
&&H=H_{0} + H_{t} + \Delta H  \nonumber\\
&&H_{0}=\epsilon_{d}\sum_{l,\alpha} n^{d}_{l\alpha}+
        \epsilon_{p}\sum_{m,\alpha}n^{p}_{m\alpha}+
        U_{d}\sum_{l}n^{d}_{l\uparrow}n^{d}_{l\downarrow}, \nonumber\\
&&H_{t}=t_{pd}\sum_{<lm>,\alpha}(d^{+}_{l\alpha}p_{m\alpha}+H.c.)
\nonumber \\
&&\phantom{H_{t}=} \mbox{}-t_{pp}\sum_{< mm'>,\alpha}
(p^{+}_{m\alpha}p_{m'\alpha}+ H.c.) \ , \\  \label{1a}
&&\Delta H=U_{p}\sum_{m} n^{p}_{m\uparrow}n^{p}_{m\downarrow}+
                 V_{pd}\sum_{<lm>,\alpha \beta }
                 n^{d}_{l\alpha }n^{p}_{m \beta },
\end{eqnarray}
in standard notation of holes at O($p$)  and Cu($d$)  sites.  The
sign convention for oxygen orbitals in $H_{t}$ \cite{Be3,Fr1} is
accepted.  Our approach to description of low-energy properties of
the above model \cite{Be3} is based on taking into account the main Coulomb
($U_{d}$) interaction exactly and  the others as perturbations.

In order to justificate this method we briefly describe here the
magnitudes of the three-band model's  parameters.  Different
experimental \cite{Esk0,Esk1,Za1}, atomic \cite{Fl1}, and band
calculations \cite{Hy2,Pi1,Fi1} show that: $U_{d}= 5\div7$ eV
\cite{Fl1}, $7\div11$ eV \cite{Hy2,Esk0}, $U_{p}=3\div8$ eV
\cite{Esk1}, $V_{pd}=0\div1.7$ eV.  $U_{p}$ is always less than $U_{d}$.
There is a general agreement for the Cu-O system that $\Delta = \epsilon
_{p}-\epsilon _{d}$ is always $>0$ and $<U_{d}$ \cite{Za1}.  It reflects the
facts that the first hole in the unit cell is predominantly at the Cu site
and the added hole has an oxygen's  character. $t_{pd}=1\div1.6$ eV (and it
is unlikely that it is less than $1$ eV) , $t_{pp}= 0.5\div0.7$ eV
\cite{Hy2,Pi1,Fi1}.  This set of magnitudes will be called hereafter the
{\it realistic region} of parameters.

The consistent low-energy reduction of the three-band model to the
generalized t-J model has been performed in previous works
\cite{Be1,Be2,Be3}.
Our method of the low-energy reduction has been  based on construction
of a set of local states with different number of holes over the filled
atomic orbitals.  The most essential states are:
\begin{enumerate}
\item The vacuum state or the vacancy which is simply
\begin{equation}
\label{2}
|v>=|0> ,
\end{equation}
\item The one-hole states which represent the ground state of CuO$_{2}$ plane
\begin{equation}
\label{2a}
|f \alpha> \equiv |\alpha> = U|d\alpha> - V|p\alpha> ,
\end{equation}
where $|d\alpha > $ and $|q\alpha>$ are the copper and symmetrical
oxygen hole states with spin projection $\alpha $, respectively.
\item The two-hole states which are the Zhang-Rice singlets:
\begin{eqnarray}
\label{2b}
|c>&=& U_{1}|d\uparrow d\downarrow>+
            V_{1}|q\uparrow q\downarrow> \nonumber \\
        & &\mbox{}+W_{1}(|d\uparrow q\downarrow>-
                  |d\downarrow q\uparrow>)/\sqrt{2} .
\end{eqnarray} \end{enumerate} The coefficients $U,V, U_{1},V_{1,}W_{1}$ are
functions of the parameters of the three-band model \cite{Be3}.  At half
filling Hamiltonian  (\ref{1},\ref{1a}) is reduced to the Heisenberg
Hamiltonian with spin-$1/2$ which are antiferromagnetically
ordered due to the second-order virtual transitions through the set of
two-hole states.  Note, that the above named spins-$1/2$ are exactly states
$|f\alpha>$ (Eq.\ (\ref{2a})).

It has been shown \cite{Be2,Be3} that for
the case near to half filling the Hamiltonian of the three-band model is
reduced to the Hamiltonian of the t-J like model of singlets, vacancies
and spins:
\begin{eqnarray}
\label{3}
H_{t-J}&=&(E_{v}-\mu)\sum_{l}X^{vv}_{l}+
      (E_{c}+\mu)\sum_{l}X^{cc}_{l} \nonumber \\
     & &\mbox{}+t_{e}\sum_{< ll'>, \alpha}X^{v\alpha}_{l'}X^{\alpha v}_{l}+
         t_{h}\sum_{< ll'>, \alpha}X^{c\alpha}_{l'}X^{\alpha c}_{l}
\nonumber \\
     & &\mbox{}+J\sum_{< ll'>}{\bf S}_{l}{\bf S}_{l'}.
\end{eqnarray}
Where $ X^{ab}_{l} \equiv |al> < lb|$ are the Hubbard
operators at the site $l$, ${\bf S}_{l} = {\mbox{\boldmath
$\sigma$}}_{\alpha\beta} X^{\alpha \beta }_{l}/2$.  The constants $E_{v}$ and
$E_{c}$ are the local energies of the vacancy and singlet; $\mu$ is the
chemical potential; $t_{e}$ and $t_{h}$ are the hopping integrals for the
vacancy and singlet (electron and hole), respectively; $J$ is the exchange
constant.   All five parameters $E_{v},E_{c},t_{e},t_{h}$ and $J$ are
functions of the three-band model parameters.  It has been shown that
relative magnitudes of the omitted terms in the Hamiltonian $H_{t-J}$ (Eq.\
(\ref{3})) are of the order of ten percent \cite{Be2,Be3}.

We believe that
the Hamiltonian (\ref{3}) describes many important properties of the
cuprates.  Therefore, the real values of its parameters are of great
interest.  As was noted above, the parameters of the primary model
 (\ref{1},\ref{1a}) are known with low precision.  In this situation
calculation of the observable quantities is an urgent issue since it provides
a way to fix parameters of Hamiltonians (\ref{1},\ref{1a}) and
(\ref{3}). The best-defined experimental values which we can describe quite
accurately are the superexchange constant $J$  and the
charge-transfer gap $E_{gap}$. The experimental values of $J$
are $0.14$ eV and $0.17$ eV for the lantan and yttrium systems, respectively
\cite{Ro,Hay1}. These values of $J$ follow from measurement of the
velocity of sound for magnons.  The value of the charge-transfer gap is known
from a variety of optical measurements
\cite{Fa1,Za1,Fi1,Uc1} and is near to $2.0$ eV.
Observation of photoconductivity at the same energies shows that the
excitations result in separated electrons and holes \cite{Fa1,Uc1}.
We have taken most of clear experimental features of the charge-transfer
spectrum from Ref. \cite{Fa1} where photoconductivity as well as reflectivity
data for La$_{2}$CuO$_{4}$ are presented.
%***************************************************************************
\section {Calculation of the observable quantities}

%***************************************************************************
The expression for the\ AF\ coupling    constant $J$ in the framework of
our approach is
\begin{equation}
\label{4}
J=-2h_{1}V^{4}U_{p}+\sum_{n}x_{n}\frac{|D_{n}|^{2}}{\Delta E_{n}}.
\end{equation}
The first term in Eq.\ (\ref{4})
represents the exchange energy between two holes (spins)
due to the repulsion at an
oxygen.  This contribution has the ferromagnetic sign and arises as an
exchange interaction between the hole states  (\ref{2a}) due to their
nonlocal nature.  The constant $V$ is defined in Eq.\ (\ref{2a}), $h_{1}$ is
in Appendix.  The second term in Eq.\ (\ref{4}) represents the correction to
the energy due to the virtual transition of hole from the state Eq.\ (\ref{2a})
into the two-hole states and back \cite{Be3}.  Here $n$ numerates the two-hole
states; matrix elements of transitions $D_{n}$ were calculated in Ref.
\cite{Be3}, $\Delta E_{n}$ are differences in energies between the energy
of the vacancy and two-hole states at neighbor sites and the ground
state energy (see Fig.\ \ref{fig1}) ; the coefficients $x_{n}=4 $ for the
singlet and $x_{n}=-2$ for triplet two-hole states.

The most general expression for the charge-transfer energy is:
\begin{equation}
\label{5}
E_{gap} = E^{N-1}_{min}-E^{N}_{g}+E^{N+1}_{min}-E^{N}_{g},
\end{equation}
where $N$ refers to the total number of electrons, $E^{N}_{g}$ is the
ground state energy, $E^{N\pm1}_{min}$ is the minimal energy of a
system with one removed and added electron.  For our system
Eq.\ (\ref{5}) can be specified as
\begin{equation}
\label{6}
E_{gap} = E^{0}_{gap}+\Delta E_{e}+\Delta E_{h} ,
\end{equation}
where $E^{0}_{gap}$ is the difference in energies between a singlet and
vacancy at local states separated by large distance and the ground
state (see Fig.\ \ref{fig2}); $\Delta E_{e}$ and $\Delta E_{h}$  are
depths of bands for electron and hole (vacancy and singlet). $E^{0}_{gap}$
can be calculated in the framework of the three-band model, whereas for the
calculation of $\Delta E_{e}$ and $\Delta E_{h}$  we will use the t-J model.

The expression for $E^{0}_{gap}$ in terms of Eq.\ (\ref{3}) is very simple:
\begin{equation}
\label{7}
E^{0}_{gap} = E_{c}+E_{v} .
\end{equation}
The values of $\Delta E_{e}$ and $\Delta E_{h}$ can be determined from
numerous analytical and numerical calculations \cite{Da1}-\cite{Su2},
\cite{Da2,Au1,Ed2,Po1,Su4}
of the dispersion relation $\epsilon ({\bf k})$ for one hole in the t-J
model on an antiferromagnetic background.  There is a general agreement that
the hole (or vacancy) on the antiferromagnetic background creates a magnetic
polaron of a small radius \cite{Lu2,Au1}, or, in the other words, the
carriers are strongly dressed by the spin waves \cite{Da2}. The influence of
antiferromagnetizm and strong correlations are manifested  in a special form
of dispersion relation $\epsilon ({\bf k})$. For our calculations we use the
results from the earlier work by Sushkov Ref. \cite{Su1} where
hole wave function
and $\epsilon ({\bf k})$ were obtained variationally:
\begin{eqnarray}
\label{8}
\epsilon ({\bf k})&=&1.32J \nonumber \\
                  & &+\frac{1}{2}\left(\tilde{\Delta}J
                  -\sqrt{\tilde{\Delta}^{2} J^{2}+16t^{2}
                     [(1+y)-(x+y)\gamma ^{2}_{\bf k}]}\ \right)
\nonumber \\
\hbox{and} \nonumber\\
\Delta E&=&1.32J+\frac{1}{2}\left(\tilde{\Delta}J
        -\sqrt{\tilde{\Delta}^{2}J^{2}+16t^{2}(1+y)}\ \right), \end{eqnarray}
where for N\`{e}el background $\tilde{\Delta} =1.33, x = 0.56, y = 0.14$.
Loss of
energy due to the broken AF bonds (four per carrier) is included. Result
for the bottom of the band at $t/J = 2.5$ from Eq.\ (\ref{8})
$\Delta E = -1.2 t$ coincides almost exactly  with the recent results of the
Green function Monte Carlo calculation by Dagotto, Nazarenko, and Boninsegni
\cite{Da2} $\Delta E = -1.255 t$.  Formula (\ref{8}) is quite good up
to $t/J \approx 5$ \cite{Su1}.

Let us discuss the characteristic values of all essential parameters which
determine the observable quantities $J$ and $E_{gap}$  (\ref{5}), (\ref{6}).
In the {\it realistic region} of
parameters of the three-band model one can easily obtain the
experimental values $J= 0.14$ eV and $J =0.17 $ eV for lantan and yttrium
systems. The value for $E^{0}_{gap}$ (\ref{7}) was obtained in Ref.
\cite{Be3} and is equal to $3.2$ eV in the same region of parameters.  The
characteristic values of $\Delta E_{e}$ and $\Delta E_{h}$ (\ref{8})
depend on
ratios $t/J$ for electrons and holes.  These ratios weakly differ and for
a typical value of $t/J = 2.5 $
we have $\Delta E_{e} \approx \Delta E_{h} \approx 0.42$
eV.  Thus, the overall gain in energy due to magnetic polaron formation is of
the order of $1$ eV, which is comparable with the experimentally observed
$E_{gap}\approx 2.0\pm0.1$ eV. Therefore, the magnetopolaron effect gives
essential contribution in the value of the charge-transfer gap.

In the paper \cite{Fa1}  it was proposed that the usual phonon polaron
effect contributes in the observable values of the charge-transfer
spectrum.  The corresponding gain in energy was estimated as $0.5$ eV.
However, we suppose that due to the Frank-Condon principle the usual
polaronic effect does not contribute in the optical transition.
Magnetopolaron effect has no such restriction since it involves
electronic degrees of freedom only.  We will return to this question in
Discussions.

%*****************************************************
\section{Parameters sensitivity}
%*****************************************************

Thus, we find superexchange constant $J$ (\ref{4}) and
charge-transfer gap
$E_{gap}$ (\ref{6}) (\cite{Esk1,Ro,PP})  as functions of the three-band
model parameters:
\begin{eqnarray}
\label{9}
&&J=J(t_{pd},t_{pp},\Delta , U_{d},U_{p},V_{pd}), \nonumber \\
&&E_{gap}=E_{gap}(t_{pd},t_{pp},\Delta ,
U_{d},U_{p},V_{pd}).  \end{eqnarray}
Both observable quantities are strongly depend on
hopping integrals and  $\Delta = \epsilon _{p} - \epsilon _{d}$, that
provides the way of fixing these latter by the experimental values of the
first.

As was discussed earlier, abundance of the parameters  makes questionable
the calculation of the effective hopping integral for the t-J
model from the three-band model  for real CuO$_{2}$  planes.  While
Coulomb repulsions are known with a fair precision ($30\%
-50\%$), the situation is complicated due to a very low
precision of the direct determination of $t_{pd},t_{pp}$ and $\Delta $ ,
which mostly affect all effective parameters.
Previously, the above  parameters have been determined from
the analysis of spectroscopic data \cite{Esk0,Esk1}.  In our recent work
we fitted $\Delta $ to experimental value of $J$ \cite{Be3}.

Now, on the basis of a better understanding of the charge-transfer process
and more accurate calculations, fixing of the worse known parameters using
experimental values of $J$ and $E_{gap}$ suggests itself.  We will
show that this procedure keeps effective $t_{h}$ inside a narrow enough
region.

Firstly, for further discussion we define O-O hopping as $t_{pp}
=\gamma t_{pd}$.  In order to characterize the above mentioned strong
dependence of the parameters (\ref{9})
one can calculate $\Delta (t_{pd})$ at fixed
$J$ or at fixed $E_{gap}$, with other parameters (Coulomb repulsions
and $\gamma $) as constants in the {\it realistic region}.
We evaluate $\Delta $
vs $t_{pd}$ at $J =140$ meV and $170$ meV, $E_{gap}=2.0$ eV and $2.5$ eV
(see Fig.\ \ref{fig3}).  Note that the profiles of the curves resemble those
at the diagram of $U(t)$  for the simple one-band Hubbard model where $E_{gap}
= U + 2W $, $W =-\alpha t $, $J =4t^{2}/U $, and the crossing point
uniquely determines $U$ and $t$.

To be more specific, we firstly determined $\Delta $ for constant $J$ at
an arbitrary $t_{pd}$ and further  moved up or down along the curve $J$
= const to fix the value  of $E_{gap}$. We used the data for La$_{2}$CuO$_{4}$
$J=140$ meV \cite{Hay1} and $E_{gap}=2.1$ eV (photoconductivity) \cite{Fa1}.
Figures\ \ref{fig4},\ref{fig5},\ref{fig6} show the parameter
of our prime interest: the effective
integral for hole in the t-J model.  Parameter $\gamma = t_{pp}/t_{pd}$
is  $ 0.5, 0.7, 0.3$ for Figs.\ \ref{fig4},\ref{fig5},\ref{fig6},
respectively. In all figures
simple dotted curve corresponds to $V_{pd}=U_{p}=0$, dotted ones with crosses
$V_{pd}=0$, $U_{p}=3$ eV, $6$ eV, dotted ones
with triangles $V_{pd}=0.5$ eV $1$ eV, $U_{p}$=0, and full curves correspond to
including both Coulombs
$V_{pd}=0.5$ eV, $U_{p}=3$ eV (upper), $V_{pd}=1$ eV, $U_{p}=6$ eV (lower).
The maximum on the first three curves is due to transition from $\Delta
>U_{d}$ (unrealistic range) to $\Delta < U_{d}$.  All variations of
$t_{h}(U_{d}, V_{pd}, U_{p},\gamma )$ actually show only weak dependence, and
in the most preferential region, when all Coulombs are included, $t_{h}$ lies
between $(2.4\div2.7)J$.  We believe, that our consideration is quite
accurate and well justified.  Hence, one can hope that the interval for $t/J$
obtained above provides the basis for {\it quantitatively}
correct calculations in
the framework of the t-J model.  For example, for the recently proposed
mechanism of superconductivity in the t-J model which provides a very
$t/J$-sensitive (exponentially) gap value \cite{Su3}.

Also some other features can be achieved.  Figures\
\ref{fig7},\ref{fig8},\ref{fig9} represent effective hopping $|t_{e}|$ for
vacancy, $t_{pd}$ and $\Delta$ respectively, vs other parameters.  Here
always $\gamma  =0.5$.  Strong support of our fixing  procedure is that
selfconsistently determined $t_{pd}$ and $\Delta $ (Figs.\
\ref{fig8},\ref{fig9}) lie in the most appropriate region.  From our
calculation $t_{pd}=1.2\div1.4$ eV, $\Delta = 2.5\div4.5$ eV , that is really
close to cluster calculation of Eskes and Sawatzky \cite{Esk1} and to the
results of other groups \cite{Hy2,Pi1,Fi1}.
%************************************************************************
\section {Excitonic  state}

%************************************************************************
The problem of consistently taking into account the Coulomb interaction
of the carriers in the framework of the t-J - like model remains
open.  Several recent works are devoted to this problem \cite{Em3}.

As was suggested earlier \cite{Em1}, the short-range part of this
Coulomb interaction may be kept by inclusion of the nearest-neighbour Cu-O
repulsion.  Since $V_{pd}$ is included in our effective model, one can
expect that some effects will be caught.

In the recent work \cite{Fa1} a  kink in the optical reflectivity
somewhere below the charge-transfer peak (at $1.75$ eV) was observed.  Since
it had no associated photoconductivity, it was related to
creation of exciton.  An essential role of the short-range Coulomb
interaction was also discussed in Ref. \cite{Fa1}. In our way of
reasoning the exciton state, if it exists, is the Frenkel exciton,
because of short-range nature of interactions.

Now it is evident that an effective attraction can result from pure
magnetopolaron effect in the t-J model \cite{Su1,Su4}.  As was
shown in works \cite{Po1,Su4}, "contact" interaction of two holes
(without charge) dressed by spin fluctuations is attractive  for a special
symmetry of the wave-function. However, the associated energy is very small
($\leq J/3 \approx 0.04$ eV).

It is possible to combine the ideas of the short-range Coulomb and
magnetopolaron effects. The difference between $E_{gap}^{0}$  (\ref{7})
when hole and electron are separated and $\Delta E_{n}$ (Fig.\ \ref{fig4})
(where $n$ denotes the lowest singlet) when they are close, is the effective
Coulomb attraction of the "bare" hole and electron (singlet and vacancy).
We write an addition to Eq.\ (\ref{3}) as
\begin{equation}
\label{10}
\Delta H_{c} =
-V_{c}\sum_{< ll'>,\alpha \beta} n^{e}_{l\alpha}n^{h}_{l'\beta},
\end{equation}
where $n^{e}$ is the electron number operator, $n^{h}$ is the
hole number operator.  We have found that $V_{c}$ is really almost
independent from $U_{d}$ and $U_{p}$ and $V_{c}\approx 0.4 V_{pd}$.  "Bare"
electron-hole attraction itself does not mean Frenkel's exciton effect.  One
has to show that "dressed" electron and hole placed closely possess the lower
energy  than the mobile ones.  In order to specify the magnetopolaronic
language  we reproduce here the wave function of the magnetic polaron from
the Ref.\cite{Su1}. For the Ising background  it has the more simple form:
\begin{eqnarray}
\label{12}
&&\psi^{+}_{\uparrow \bf k}=\frac{1}{\sqrt{N/2}}
\sum_{n} d^{+}_{n \uparrow}\exp (i{\bf k}{\bf r}_{n}), \nonumber \\
&&d^{+}_{n \uparrow}=\nu h^{+}_{n \uparrow}+\mu S^{+}_{n} \sum_{n' \in < n>}
h^{+}_{n'\downarrow},
\end{eqnarray}
where $n\in$
sublattice with the spin $s =-1/2 $, $h^{+}_{n\alpha }$ primarily hole
operators;  at $t/J >1$, $\nu ^{2}\approx 1/2 $, $\mu ^{2}\approx 1/8$.
Thus, ansatz consist of mixture of "bare" hole and holes with one overturned
spin.  Contact interaction of these polarons with opposite spins was
considered in Ref.\cite{Su4} (see Fig.\ \ref{fig10}).  The gain in energy
for "attracting polarons" arises from the pure t-J model effects \cite{Su4}
and effective Coulomb attraction (\ref{10}).  The loss in energy is from
restriction of the mobility.  Competition of these evident effects (without
Coulomb interaction) provides bound states up to $t/J =2\div3$
\cite{Po1,Su4}. Simply acting in a spirit of magnetic polaron
interaction \cite{Su4} we obtain $\Delta E_{exc}\approx 0.35 V_{c}$. Thus, at
$V_{pd} = 1$ eV, $\Delta E_{exc} = 0.14 $ eV.

More accurate variational construction of the exciton-magnetopolaron
wave-function (on the  Ising background) yields $\Delta
E_{exc}\approx 0.5 V_{c}$ (at $t/J = 2.5\div3$).
This wave-function consists of the mixture of "bare" hole and electron
at neighbor sites and hole and electron with overturned spins.
It is schematically shown in Fig. \ref{fig11}a.  Thus, at
$ V_{pd}=1$ eV, $\Delta E_{exc}\simeq 0.2 $ eV, that is slightly  less than
the observed $\Delta E^{exp}_{exc}= 0.25\div0.35 $ eV \cite{Fa1}.
In our calculations the interaction between next-nearest-neighbor
magnetic polaron was neglected, which produced small effect for the pure
t-J model \cite{Su4} but may be essential for the problem with
attraction.  (These configurations are shown in Figs.\
\ref{fig11}b,c).  Also, the answer may partly lie in the rest of
the long-range Coulomb interaction.
%****************************************************************************
\section {Discussions}

%****************************************************************************
The detailed quantitative consideration of some of the effective
parameters of the low-energy models related to description of the high
- $T_{c}$ superconductors presented in this work relies heavily on our
earlier works.  In these works consistent mapping of the three-band
Hubbard model onto the  effective t-J model
(\cite{Be1,Be2,Be3}) has been
produced. Taking into account all essential interactions enables us
to correctly calculate local energies of various set of states with
different number of particles and matrix elements of interesting
transitions.  Combination of properties of the local "bare" hole and
electron (ZR-singlet and vacancy) and their magnetic polaron nature
as the carriers allows us to approach the calculation
of some observable quantities adequately.

We have calculated the superexchange $J$ and
charge-transfer gap $E_{gap}$.  Their experimental values strongly constrict a
possible
variation interval for the quantity of great interest: the $t/J$ ratio
in the t-J model.
Selfconsistent calculation of this ratio for a wide range of
parameters places it into the region $t/J = 2.4\div2.7 $.  Narrowed ranges
for the three-band model parameters have also been determined:
$t_{pd}=1.2\div1.4$ eV, $\Delta = 2.5\div4.5 $ eV. They coincide quite well
with earlier cluster calculations, that supports our selfconsistent
procedure.  An excitonic state of the Frenkel type induced by the short-range
Coulomb interaction with the energy lower than the charge-transfer transition
approximately by $0.2 $ eV is found.

We have also compared the width of the
peak in the $\epsilon _{2}(\omega )$ at $2.3 $ eV from Ref. \cite{Fa1} which is
of
the order of $0.5 $ eV with the total width of the charge-transfer spectrum.
This total width is equal to the combined width of the  vacancy and singlet
bands.  According to Eq.\ (\ref {8}), the
width of the hole band is $W_{h} =2.0 J$ at $t_{h}=2.55J $ and width of the
electron band is $W_{e} = 2.2J $ at $t_{e}=2.75 J$. Resulting total
width of the charge-transfer spectrum is about $0.6$ eV.
Thus, the narrowness of the $\epsilon_2(\omega)$ spectrum can also be easily
reproduced by magnetic polaron language.

One of the essential questions for the CuO$_{2}$ - planes systems, which
we only briefly touched, is the phonon polaron effect.  If it does not
have a projection on optics, or, as we believe, the Frank-Condon principle is
applicable, our calculated $t_{h}/J$ ratio is the upper limit of the
real parameter.  This is due to mass renormalization for real carriers.
There is another view on the polaron effect (see Ref. \cite{Fa1,Ch1}).
It is stated that electron-phonon interaction lies in an intermediate
range and thus the Frank-Condon principle is not obeyed.  If such is the
situation, we underestimate the depth of the bands (or
overestimate $E_{gap}$), and effective $t_{h}$ should be increased.
Our estimation shows that this increasing of $t_h$ is no more than
30\%.  Naturally, this problem requires additional
investigations.
\vskip 1.cm

%***********************************************************************
{\Large \bf Acknowledgments} \\ \vskip 0.1cm
%***********************************************************************
This work was supported partly by the Council on Superconductivity of
Russian Academy of Sciences, Grant No 93197; Russian Foundation for
Fundamental Researches, Grant No 94-02-03235;
The Competition Center for Natural Sciences at St.-Peterburg State
University, Grant No 94-5.1-1060;
by the Scientific-Technical Program "High - Temperature Superconductivity"
as part of the state program "Universities as Centers for Fundamental
Research".
%*****************************************************************

\appendix
\section*{}

In this Appendix we present some details of the technical treatment of the
problems discussed  in the paper.  According to our previous work
\cite{Be3},  we use the following  transformation from the primary
oxygen $p_{{\bf l}x},p_{{\bf l}y}$ operators to the operators $q_{\bf
l}, \tilde q_{\bf l}$  of the symmetrical and antisymmetrical oxygen states:
\begin{eqnarray}
\label{A1}
(q_{\bf l}, \tilde q_{\bf l})&=&\sum_{\bf k}[p_{{\bf k}a}\cos(k_{x}/2)\pm
p_{{\bf k}b}\cos(k_{y}/2)] \nonumber \\
                           & &\mbox{}\times (1+\gamma _{\bf k})^{-1/2}
                              \exp(i\bf kl),
\end{eqnarray}
where $p_{{\bf k} a}$ is the Fourier image of $p_{{\bf l}x} $ for
$q_{\bf l}$ and $p_{{\bf l}y}$ for $\tilde q_{\bf l}$; $p_{{\bf k}b}$ is
the Fourier image of $p_{{\bf l}y}$ for $q_{\bf l}$ and $p_{{\bf l}x}$ for
$\tilde q_{\bf l}$; $\gamma_{\bf k}$=(cos(k$_{x}$a)+cos(k$_{y}$a))/2.  The
summation in Eq.\ (\ref{A1}) is produced over the Brillouin zone, and the
lattice constant a$=1$.

Since the groundstates of both undoped and doped systems do not consist
of antisymmetrical oxygen state \cite{Esk0,Esk1,Be3} , the
reformulated Hamiltonian (\ref{1},\ref{1a}) where only essential degrees
of freedom are kept, is conveniently expressed through the local and hopping
parts \cite{Be3}. The local part is:
\begin{eqnarray}
\label{A2}
&&H_{loc}=\epsilon_{d}\sum_{l,\alpha} n^{d}_{l\alpha}+
(\epsilon_{p}-\mu_{0}t_{pp})\sum_{l,\alpha}n^{q}_{l\alpha}\nonumber \\
&&\phantom{H_{loc}=}\mbox{}+U_{d}\sum_{l}n^{d}_{l\uparrow}n^{d}_{l\downarrow}
   +V_{pd}f_{0}\sum_{l,\alpha \beta}n^{d}_{l\alpha}n^{q}_{l\beta} \nonumber \\
&&\phantom{H_{loc}=}\mbox{}+U_{p}h_{0}\sum_{l}n^{q}_{l\uparrow}n^{q}_{l\downarrow}
   +2t_{pd}\lambda_{0}\sum_{l,\alpha}(d^{+}_{l\alpha}q_{l\alpha}+ H.c.)\ ,
\nonumber\\
&&\Delta H_{int}=V_{pd}f_{1}\sum_{< ll' >, \alpha \beta}
          n^{d}_{l\alpha}n^{q}_{l'\beta}\nonumber \\
&&\phantom{\Delta H_{int}=}\mbox{}-2U_{p}h_{1}\sum_{<ll'>}
        ({\bf S}^{q}_{l}{\bf S}^{q}_{l'}-\frac{1}{4}n^{q}_{l}n^{q}_{l'})\ ,\\
&&\mbox{with  } {\bf S}^{q}_{l}=\frac{1}{2}q^{+}_{l\alpha}
{\mbox{\boldmath $\sigma$}}_ { \alpha \beta}q_{l\beta},
\ \ \ \ n^{q}=n^{q}_{l\uparrow}+n^{q}_{l\downarrow}.
\end{eqnarray}
The hopping part is:
\begin{eqnarray}
\label{A3}
&&H_{hop}=2t_{pd}\lambda_{1}\sum_{<ll'>, \alpha}
(d^{+}_{l\alpha}q_{l'\alpha}+H.c.) \nonumber \\
&&\phantom{H_{hop}=}\mbox{}-2t_{pp}\mu_{1}\sum_{<ll'>, \alpha}
          q^{+}_{l\alpha}q_{l'\alpha} ,
\nonumber \\
&&\Delta H_{hop}=V_{pd}f'\sum_{<ll'>, \alpha \beta}
n^{d}_{l\alpha}[q^{+}_{l\beta}q_{l'\beta}+H.c.] \nonumber \\
&&\phantom{\Delta H_{hop}=}\mbox{}+U_{p}h'\sum_{<ll'>, \alpha}
                 n^{q}_{l\alpha}[q^{+}_{l\bar{\alpha}}q_{l'\bar{\alpha}}+H.c.].
\end{eqnarray}
All constants $\lambda ,\mu , f, h $ in
Eqs.\ (\ref{A2},\ref{A3}) are of Wannier
nature. Their Fourier images and magnitudes are given in Ref.\cite{Be3}.
In order to group them together we reproduce
\vskip 0.5 cm
\begin{center}
\begin{tabular}{cccc}
$\lambda _{0}$=0.9581 & $\lambda _{1}$=0.1401 & $\mu_{0}$=1.4567 &
$\mu_{1}$=0.2678 \\ $f_{0}$=0.9180       & $f_{1}$=0.2430        &
$h_{0}$=0.211 & $h_{1}$=0.059 \\ $f'$=0.1342          & $h'$=0.030 .
& &
\end{tabular}
\end{center}
\vskip 0.5cm
We have treated the
Hamiltonian $H_{loc} + \Delta H_{int}$  (\ref{A2}) in the selfconsistent
mean-field approximation \cite{Be3} that enables us to solve the problem of
the local states at site with different number of holes.  The matrix elements
of the Hamiltonian (\ref{A3}) between states with a singlet or vacancy at
different sites in initial and final states lead to the following expression
for the hopping constants:
\begin{eqnarray}
\label{A5}
&&t_{h}=2t\lambda _{1}(W_{1}V^{\prime}-\sqrt{2}U_{1}U^{\prime})
        (W_{1}U^{\prime}-\sqrt{2}V_{1}V^{\prime})
\nonumber \\
&&\phantom{t_{h}=}\mbox{}+t_{p}\mu _{1}(W_{1}U^{\prime}-
     \sqrt{2}V_{1}V^{\prime})^{2}/2 \nonumber \\
&&\phantom{t_{h}=}\mbox{}-(V_{pd}f_{1}W_{1}U^{\prime}-
     \sqrt{2}U_{p}h_{1}V_{1}V^{\prime})(W_{1}U^{\prime}-
     \sqrt{2}V_{1}V^{\prime}),
\nonumber \\
&&t_{e}=\mbox{}-4t\lambda _{1}U^{\prime\prime}V^{\prime\prime}-
            t_{p}\mu _{1}(V^{\prime\prime})^{2},
\end{eqnarray}
where $U^{\prime},V^{\prime}$ are the coefficients of $|f>$ - state nearest
to singlet, $U^{\prime\prime}, V^{\prime\prime}$ - those nearest to vacancy.
The coefficients $U^{\prime},V^{\prime}$ and $U^{\prime\prime},
V^{\prime\prime}$ are slightly different from the ones in Eq.\ (\ref{2a})
due to a distortion of $|f>$ - states by the nearest vacancy or singlet.
This distortion has its origin in the
short-range Coulomb repulsions due to $U_{p}$ and $V_{pd}$ terms in the
Hamiltonian (\ref{1a}).  Note that it changes the energy of the
$|f>$ - states nearest to a vacancy or singlet.  We take this effect
into consideration when we calculate the quantity $E^{0}_{gap}$. Thus,
we have for the total energies $E_{c}$ and $E_{v}$ for a singlet and a
vacancy Eq.\ (\ref{7})
\begin{eqnarray}
\label{A6}
&&E_{c}=E_{s}+4E_{fs}-5E_{f}, \nonumber \\
&&E_{v}=E_{0}+4E_{f0}-5E_{f},
\end{eqnarray}
where $E_{s}$ and $E_{0}$ are the local energies of singlet and
vacancy;  $E_{fs}$ and $E_{f0}$ are the energies of the $|f>$ - states
nearest to a singlet or vacancy;  $E_{f}$ is the energy of $|f>$ - state
in an undoped sample.  These states are schematically shown in Fig.\
\ref{fig2}.

When the vacancy and the two-hole state are created at the neighbour
sites (see Fig.\ \ref{fig1}) the difference in energy between this state and
the ground state is determined by the relation
\begin{equation}
\label{A7}
\Delta E_{n}=E_{n}+E_{0}+3E_{fs}+3E_{f0}-8E_{f}.
\end{equation}
These energies $\Delta E_{n}$ are involved in calculation of the
superexchange constant $J$ (Eq.\ (\ref{4})) and in the energy of the Coulomb
attraction of a singlet and vacancy in Sec. V.

\begin{figure}
\caption{The nearest-neighbor two-hole state and vacancy. Black circle
denotes two-hole state, empty circle denotes vacancy. Crosses are
one hole states (spins). }
\label{fig1}
\end{figure}
\begin{figure}
\caption{The separated ZR singlet (hole) and vacancy (electron). Black
       circle denotes singlet, empty circle denotes vacancy. Crosses are
       one-hole states (spins).}
\label{fig2}
\end{figure}
\begin{figure}
\caption{$\Delta$ vs $t_{pd}$ at constant $J$ or $E_{gap}$.
       $U_{d}=7$ eV, $U_{p}=3$ eV, $V_{pd}=1$ eV, $t_{pp}/t_{pd}=0.5$.
       Full curve - $J=140$ meV, dashed curve - $J=170$ meV, full curve
       with markers - $E_{gap}=2.0$ eV, dotted curve - $E_{gap}=2.05$ eV.}
\label{fig3}
\end{figure}
\begin{figure}
\caption{Effective hopping integral for t-J model hole vs $U_{d}$.
       Dotted line - $V_{pd}=U_{p}=0$,
       dotted with crosses - $V_{pd}=0$, $U_{p}=3$, $6$ eV,
       dotted with triangles - $U_{p}=0, V_{pd}=0.5$, $1$ eV,
       full curves - $V_{pd}=0.5$ eV, $U_{p}=3$ eV (upper),
                   $V_{pd}=1$ eV, $U_{p}=6$ eV (lower), $\gamma =0.5$.}
\label{fig4}
\end{figure}
\begin{figure}
\caption{All notations as for Fig.4; $\gamma =0.7$.}
\label{fig5}
\end{figure}
\begin{figure}
\caption{All notations as for Fig.4; $\gamma =0.3$.}
\label{fig6}
\end{figure}
\begin{figure}
\caption{Effective hopping integral for electron in the t-J model vs
$U_{d}$, curves markers as for Fig.4; $\gamma =0.5$.}
\label{fig7}
\end{figure}
\begin{figure}
\caption{Cu-O hopping integral vs
$U_{d}$, curves markers as for Fig.4; $\gamma =0.5$.}
\label{fig8}
\end{figure}
\begin{figure}
\caption{$\Delta$ vs $U_{d}$; $\gamma =0.5$.}
\label{fig9}
\end{figure}
\begin{figure}
\caption{Configuration of interacting magnetic polarons from Ref.\ [36].}
\label{fig10}
\end{figure}

\begin{figure}
\caption{(a) Exciton-magnetopolaron; (b),(c) next-nearest-neighbor magnetic
polarons.}
\label{fig11}
\end{figure}

\end{document}